\documentclass[10pt,aps,prd,noshowpacs,superscriptaddress,notitlepage,nofootinbib]{revtex4-1}
\usepackage{amssymb}
\usepackage{graphicx,verbatim}
\usepackage{epstopdf}
\usepackage{slashed}
\usepackage{comment}
\begin{document}

\title{ The QCD running charge and its RGI three-gluon vertex parent in the Pinch Technique  }
\author{ John M. Cornwall}
\affiliation{Department of Physics and Astronomy, University of California, Los Angeles CA 90095}
\email[Email]{cornwall@physics.ucla.edu}

\begin{center}
Talk given at the symposium ``Quantum Chromodynamics:  History and Prospects", Oberw\"olz, Austria, September 2012
\end{center}

\begin{abstract}

We give a brief review of an elementary extension of the Pinch Technique (PT) that yields renormalization-group invariant (RGI) Green's functions.  These are also gauge- and process-independent, and show dimensional transmutation.  These are non-perturbatively IR-finite because the gluon is massive, but here we bypass all issues of actually calculating that mass.    The ghost-free Ward identities of the PT lead to the conclusion that if the dressed propagators and vertices going into the one-dressed-loop skeleton graphs for three- and four-point vertices are approximated by free vertices and free propagators with a non-running mass $m$, the one-loop output,  after some tweaking of seagulls, is   PT-RGI.  The PT-RGI gluon propagator is recovered by using the Ward identity, and turns out to be identical to an older proposal made on different grounds. Without  approximation one can decompose the all-order three-gluon vertex as $G_{\alpha\mu\nu}(p,q,-p-q)=G^0_{\alpha\mu\nu}(p,q,-p-q)G(p,q,-p-q)+\dots$ where $G^0_{\alpha\mu\nu}$ is the bare vertex and show that $G(0,q,-q)$ is equal to $\bar{g}^{-2}(q)$, the inverse square of the running charge.  This three-point parent of the usual running charge should lead to more flexible and physically-sensible phenomenology than is possible just with the running charge.  We give an explicit illustration based on an asymptotically-free theory having no spin complications, $\phi^3_6$, yielding results quite similar to those proposed long ago on phenomenological grounds for QCD.  Finally, we discuss finding the beta-function from the running charge, showing how non-perturbative terms arise, as well as possible implications of skeleton graphs with more than one loop, for the beta-function and non-leading UV behavior.
\end{abstract}

\maketitle

\section{Introduction}

The running charge plays a particularly distinguished role in QCD, both in the history of QCD and in extensive applications.  Its perturbative behavior in the UV exemplifies asymptotic freedom, and it is constantly invoked in phenomenology.  But it has its difficulties:  One must choose a single momentum scale for the running charge, while for many processes (for example, octet $q\bar{q}$ annihilation into two gluon jets)  it may be more appropriate physically to use the parent three-point function with its three momentum scales.      

The PT \cite{cornbinpap}  offers a way to construct off-shell Green's functions of a non-Abelian gauge theory (NAGT) that are gauge- and process-independent.  Recently, a trivial extension was introduced \cite{corn141} that made PT Green's functions also renormalization-group independent (RGI), meaning independent of any renormalization mass $\mu$.  This is possible because the PT leads to ghost-free Ward identities, like those of QED, and through these to the equality of certain renormalization constants.  

First, we describe conventional renormalization in order to contrast it with the PT-RGI procedure.  Consider a renormalizable\footnote{Not superrenormalizable---so either in $d=6$, or in $d=4$ with additional four-point couplings.} field theory with a bare coupling term $g_0\phi_1\phi_2\phi_3$, yielding an unrenormalized vertex $\Gamma_U$.  Use subscripts $U$ to denote fully-dressed but unrenormalized Green's functions or fields, and $R$ to denote renormalized quantities.    These are related by renormalization constants:
\begin{equation}
\label{genvert}
\phi_{Ui}=Z_i^{1/2}\phi_{Ri},\;\;g_0=\frac{Z_Vg_R}{(Z_1Z_2Z_3)^{1/2}},\;\;\Gamma_U=\frac{\Gamma_R}{Z_V}. 
\end{equation}
In this generic theory the renormalized coupling is arbitrary, and there are four different renormalization constants.  A typical renormalization equation applies to the propagators $D_i$:
\begin{equation}
\label{typrenorm}
D_{Ui}(p;\Lambda_{UV})=Z_i(\mu,\Lambda_{UV})D_{Ri}(p; \mu)
\end{equation}
in which $\Lambda_{UV}$ is a UV cutoff and $\mu$ is an essentially arbitrary mass scale---the renormalization point.  (Of course, one can introduce many renormalization points, but usually one is enough.)   Now $D_{Ri}$ is finite and independent of $\Lambda_{UV}$, but the price paid is that this propagator depends on $\mu$.  Various choices of $\mu$ are related by the renormalization group.

Our objective is to modify the PT so that this $\mu$-dependence goes away at the level of off-shell Green's functions, and not just at the level of the S-matrix.
Although we will not carry out any calculations in a ghost-free gauge such as $n\cdot A =0$, it simplifies understanding of the PT-RGI connection to imagine  using such a gauge\footnote{As in the original Pinch Technique \cite{corn076}.} for conventional (gauge-dependent) Feynman graphs.  This cannot affect any properties of the PT itself, which always deals with gauge-invariant quantities constructed from re-organizing the Feynman graphs.  We begin with pure gluonic Green's functions; it is a simple matter to add closed quarks loops,\footnote{And also closed ghost loops in covariant gauges.  The point is that such closed loops only depend on the product $\Gamma S$, where $\Gamma$ is, for example, the quark-gluon vertex and $S$ the quark propagator; this product is RGI in the PT.} which do not change any of our conclusions for Green's functions with only gluon external legs.
An essential feature of the PT is that even for an NAGT the fundamental Ward identities are QED-like.  Indeed, for the NAGT in a ghost-free gauge they are QED-like even before applying the PT re-organization.  This means, as we show below, that there is only one renormalization constant in a ghost-free gauge for the PT:  The propagator, three-vertex, and four-vertex renormalization constants are equal to a common value $Z$.  Let us denote the unrenormalized PT propagator by $d_{U\mu\nu}(p)$, the PT three-vertex by $\Gamma_{U\mu\nu\alpha}(p_1,p_2,p_3)$, and the PT four-vertex by $\Gamma^{(4)}_{U\mu\nu\alpha\beta}(p_1,p_2,p_3,p_4)$ (suppressing group indices; often we also suppress vector indices when there can be no confusion).
For future reference, we introduce the proper self-energy of the PT propagator $d_{U\mu\nu}(p)$:
\begin{equation}
\label{ptpropform}
d^{-1}_{U\mu\nu}(p)=d^{-1}_{0\mu\nu}(p)+\Pi_{U\mu\nu}(p)
\end{equation}
where $d^{-1}_{0\mu\nu}(p)$ is the inverse free propagator in whatever gauge is chosen; it carries all the gauge dependence.  The self-energy is explicitly free of all gauge dependence.  

We need not indicate whether the Green's functions in the Ward identities are unrenormalized or renormalized.  In the PT $\Pi_{\mu\nu}$ is independent of the gauge chosen, and in a ghost-free gauge it cannot depend on the vector $n_{\mu}$ that sets the gauge $n\cdot A =0$.  Therefore it is of the form
\begin{equation}
\label{piform}
\Pi_{\mu\nu}= P_{\mu\nu}(p)\Pi (p)
\end{equation}
where
\begin{equation}
\label{proj}
P_{\alpha\beta}(p) = \delta_{\alpha\beta}-\frac{p_{\alpha}p_{\beta}}{p^2}
\end{equation} 
is a projector.  [In the light-cone gauge, the self-energy does depend on $n_{\mu}$ before applying the PT, and there is another conserved form that is eliminated by the PT; see \cite{corn076}.]
Just as in QED the Ward identity implies that $Z_1=Z_2$, the present Ward identities tell us that there is just one renormalization constant $Z$, such that:
\begin{equation}
\label{renormdef}
   d_U=Zd_R;\;\;\Gamma_U=\frac{\Gamma_R}{Z};\;\;\Gamma_U^{(4)}=\frac{\Gamma_R^{(4)}}{Z};\;\; g_0^2=\frac{g_R^2}{Z}.
\end{equation} 

Throughout this paper we work in a Euclidean metric.

\section{PT-RGI Green's functions and Ward identities}

All that is necessary to form PT-RGI Green's functions is to divide the
  (proper) PT Green's functions by $g_0^2$.  The resulting Green's functions $\Delta, G$, and $G^{(4)}$ are the same whether unrenormalized or renormalized and therefore independent of the renormalization point $\mu$, as eqn.~(\ref{renormdef}) shows:  
\begin{eqnarray}
    \Delta_R =  \Delta_U & = & g_0^2d_U = g_R^2d_R, \\ \nonumber
 G_R(p_i)  =  G_U(p_i) & = & \frac{\Gamma_U(p_i)}{g_0^2}=\frac{\Gamma_R(p_i)}{g_R^2}, \\ \nonumber
  G^{(4)}_R(p_i)= G^{(4)}_R(p_i) & = & \frac{\Gamma^{(4)}_U(p_i)}{g_0^2} = \frac{\Gamma^{(4)}_R(p_i)}{g_R^2}
\end{eqnarray}

   We sometimes write 
\begin{equation}
\label{znotation}
d_U(p)=\frac{H(p)}{\tilde{Z}_U(p)},\;\;\bar{g}^2(p)=\frac{g^2}{\tilde{Z}_R(p)},\;\;
\tilde{Z}_U(p)=\frac{\tilde{Z}_R(p)}{Z}
\end{equation}
with $H_U=H_R\equiv H$ an RGI function.
In perturbation theory, $H=1/p^2$, so we  {\bf define}
\begin{equation}
\label{hdefinition}
H=\frac{1}{p^2+m^2(p^2)}
\end{equation}
at all momenta.  
Moreover, 
\begin{equation}
\label{tildez}
\tilde{Z}(p)/g^2 =\bar{g}^{-2}(p)
\end{equation}
is RGI.  Finally, we have:
\begin{equation}
\label{hruneqn}
\Delta_R(p^2)=H(p^2)\bar{g}^2(p^2).  
\end{equation}

From now on we  drop the irrelevant subscripts $U,R$ on PT-RGI Green's functions.  These relations mean that, for example, $\Delta$ is independent of $\mu$ as well as of $\Lambda_{UV}$.  We show below that the ghost-free Ward identities continue to be satisfied by the PT-RGI Green's functions.  And finally, the process of dividing by $g_0^2$ is unnecessary for all Green's functions with five or more legs, since the skeleton graphs for these are finite.

Now we come to the essence of constructing PT-RGI Green's functions.  
The PT Ward identities are:
\begin{equation}
\label{wardiden1}
p_{\mu}\Pi_{\mu\nu}(p)=0;
\end{equation}
\begin{equation}
\label{wardiden3}
   p_{1\alpha}\Gamma_{\alpha\beta\gamma}(p_1,p_2,p_3) =
   d^{-1}(p_2)_{\beta\gamma}(p_2)-d^{-1}(p_3)_{\beta\gamma}(p_3);    
\end{equation}
and 
\begin{eqnarray}
\label{wardiden4}
p_{1\mu}\Gamma^{(4)abcd}_{\mu\nu\alpha\beta} & = & f^{aeb}\Gamma^{ecd}_{\nu\alpha\beta}(p_1+p_2,p_3,p_4)\\ \nonumber
~ & + & f^{aed}\Gamma^{edb}_{\nu\alpha\beta}(p_1+p_3,p_4,p_2)\\ \nonumber
~ & + & f^{aec}\Gamma^{ebd}_{\alpha\nu\beta}(p_1+p_4,p_2,p_3).
\end{eqnarray}
Note that the Ward identities are of the same form in terms of PT-RGI Green's functions, for example:
\begin{equation}
\label{genwi}
p_{1\alpha}G_{\alpha\beta\gamma}(p_1,p_2,p_3)=P_{\beta\gamma}(p_2)\Delta^{-1}(p_2)
-P_{\beta\gamma}(p_3)\Delta^{-1}(p_3).
\end{equation} 

Certain quantities require special treatment, notably seagull graphs.  These are indeed formally PT-RGI as one can easily check, but still naively divergent, and there is no mass term in the Lagrangian to absorb this divergence.  But it has been argued \cite{papag} that, due to an equation given in \cite{corn076}, there is a cancellation between the seagull and other graphs that removes this divergence.   

\section{Schwinger-Dyson equations}

The SDEs of NAGTs are complex in any formulation, and have additional features---both complexities and simplifications---in the PT.  One key point for the PT is that it has ghost-free Ward identities allowing precise construction of the two-point function (propagator) from the three-point vertex.\footnote{In contrast to the gauge technique \cite{cornbinpap}, which attempts to approximate the three-vertex from the propagator, and which unlike the PT-RGI approach is not exact in the UV.}  (Similarly, the three-vertex can be constructed from the four-vertex, but we will not mention the four-vertex further in this initial investigation.)  Of course, the three-vertex depends on the propagator, so this sounds like a vicious circle.  We suggest here some methods for breaking into this circle.

The skeleton-graph expansion equivalent to the SDEs will be expressed in terms of fully-dressed vertices, and then the Ward identity will produce a propagator SDE with only fully-dressed vertices.  This is the easiest way to see full Bose symmetry, etc, and it also avoids difficult issues of multiplying things by infinite Zs, instead of subtracting, as in eqn.~(\ref{approxvert}) below.  Such an expansion with only dressed vertices and propagators differs in structure, but is equivalent to, the more conventional SDE treatment with some vertices being bare.

A key point of this paper is the suggestion that a good approximation to the fully-dressed one-loop SDEs is to use free vertices and free massive propagators in the loop and then apply the PT as in \cite{corn099}.   Some minor ``by hand" adjustments to seagulls yields an output  which satisfies all the requirements for a one-loop PT-RGI three-vertex and, through the Ward identity, a one-loop PT-RGI gluon propagator, both of which are well-behaved in the IR. 

\subsection{The SDE in the UV}

It turns out that the Ward identities allow a precise characterization of the UV behavior of the full non-linear SDEs; to no one's surprise, this is exactly the same as in perturbation theory, because the gauge theory is asymptotically-free.

Write the vertex as
\begin{equation}
\label{vertequ}
G_{\alpha\beta\gamma}(p_i)=G^0_{\alpha\beta\gamma}G(p_i)+\dots
\end{equation}
where
\begin{equation}
\label{bornvert}
G^0_{\alpha\beta\gamma}(p_i)=(p_1-p_2)_{\gamma}\delta_{\alpha\beta}+(p_2-p_3)_{\alpha}\delta_{\beta\gamma}
+(p_3-p_1)_{\beta}\delta_{\alpha\gamma}
\end{equation}
is the Born term.
For the propagator, write
\begin{equation}
\label{zeromom}
\Delta_{\beta\gamma}(q)  = \Delta (q)P_{\beta\gamma}(q)+ \mathrm{gauge-fixing \;term}
\end{equation}
where $q$ is one of the $p_i$.
If we save only the coefficient $G$ of the Born term in the vertex, the Ward identity has two terms, each of which says:
\begin{equation}
\label{uvwi}
G(p_i)\Delta (q)=\frac{1}{q^2}.
\end{equation}
Clearly, this cannot be an equation for all $p_i$, but in the UV region $p_i\sim q$ with $q$ large, it is true to leading order of logarithms; non-leading orders are compensated in the terms we omitted in the vertex.\footnote{Note that omitted terms are necessarily non-leading and cannot contribute to UV divergences; if they did, there would have to be a corresponding counterterm in the Lagrangian.}   We do not have space here to show a crucial result, which is however elementary for $\phi^3_6$:  Using a ghost-free gauge for the PT, all one-loop skeleton graphs for the vertex depend only on the product $G\Delta$.  But this in the UV is the same, according to eqn.~(\ref{uvwi}), as if all vertices and propagators are free.\footnote{Since the one-loop skeleton graphs contain four-vertices, it is necessary to show that this result holds in the presence of such four-vertices, which we have done.}

\subsection{The SDE in the IR}

\begin{figure}
\begin{center}
\includegraphics[width=5in]{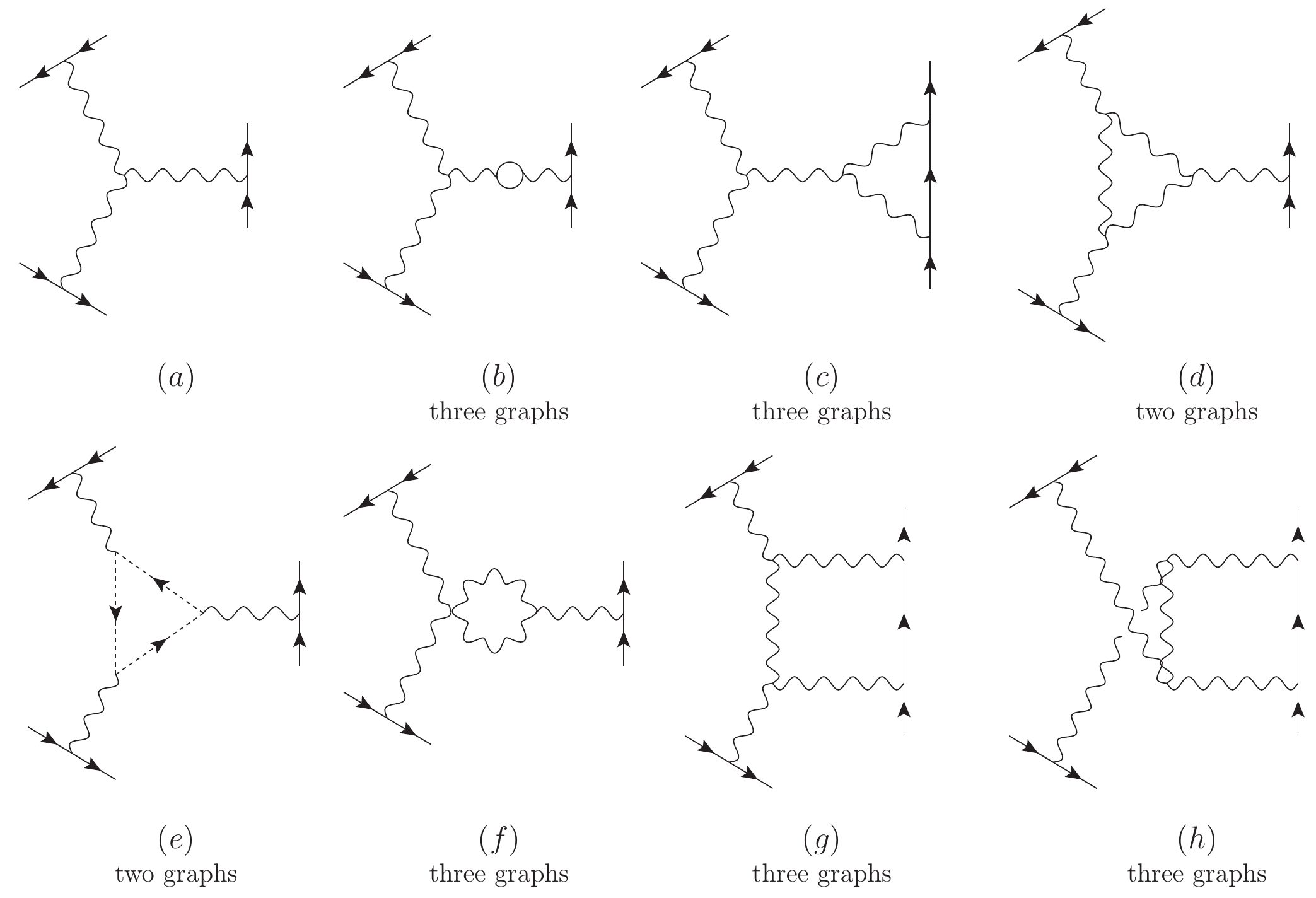}
\caption{\label{3gv} The PT graphs for the one-loop three-vertex are extracted from an S-matrix element for on-shell quarks through pinching out internal quark lines with Ward identities, as explained in \cite{cornbinpap}.}
\end{center}
\end{figure}

What happens beyond the UV and into the IR, where eqn.~(\ref{uvwi}) needs correction?  This is a hard problem, and we have only some suggestive remarks.  These are based on extending the calculation in \cite{corn099} of the perturbative one-loop three-vertex (see Fig.\ref{3gv}) to be phenomenologically useful in the IR. 
We are interested in these graphs as skeleton graphs, so all vertices and lines are to be thought of as dressed.
The simplest possible IR stabilization of an asymptotically-free theory is to change free massless propagators into massive ones in the PT calculation of the one-loop three-vertex of \cite{corn099}.  This calculation is effectively in the background-field Feynman gauge, so we use as the propagator
\begin{equation}
\label{proppart}
d_{\alpha\beta}(p)=P_{\alpha\beta}(p) d(p)+\frac{p_{\alpha}p_{\beta}}{p^4} 
\end{equation}
and
\begin{equation}
\label{invproppart}
d^{-1}_{\alpha\beta}(p)=P_{\alpha\beta}(p)d^{-1}(p)+p_{\alpha}p_{\beta} 
\end{equation}
 with $d(p)=1/(p^2+m^2)$.  It is important for what follows that $m^2$ is constant in momentum space.  The corresponding vertex is the free vertex in the background-field Feynman gauge, which is not Bose-symmetric on all three gluon lines, but obeys a simple Ward identity on one of the lines.  We always choose this special line to be one of the gluon lines attached to quark lines in Fig.~\ref{3gv}.  The Ward identity is:
\begin{equation}
\label{usuwi}
p_{1\alpha}\Gamma^F_{\alpha\mu\nu}(p_1,p_2,p_3)=\delta_{\mu\nu}[(p_3^2+m^2)-(p_2^2+m^2)]=
\Delta^{-1}_{\mu\nu}(p_3)-\Delta^{-1}_{\mu\nu}(p_2).
\end{equation} 
and the crucial point is that if $m^2$ is constant in momentum space, the mass terms in the inverse propagators cancel in this Ward identity.  In consequence, since the PT is based on repeatedly applying such Ward identities, one can choose all numerators in the graphs of Fig.~\ref{3gv} to be the same as in the massless case.  

We are not trying to calculate $m^2$ self-consistently, which would involve a careful treatment of seagulls, so we take the liberty of defining seagull terms, with numerators $\sim m^2$, as necessary for satisfying Ward identities.  Furthermore, for the same reason,
at the end of the calculation we must add a term as prescribed in \cite{corn090} to account for the massless poles in the propagator side of the Ward identity.  This term cannot contribute to S-matrix elements, since it annihilates conserved currents, and is of the form:
\begin{eqnarray}
\label{v2def}     
    V_{\alpha\beta\gamma}(p_1,k_3,-k_2) & = & \frac{m^2}{2}[-\frac{p_{1\alpha}k_{3\beta}(p_1-k_3)_{\lambda}P_{\lambda\gamma}(k_2)}{p_1^2k_3^2} +\\ \nonumber
& & \frac{k_{2\gamma}k_{3\beta}(k_3+k_2)_{\lambda}P_{\lambda\alpha}(p_1)}{k_2^2k_3^2}-
\frac{p_{1\alpha}k_{2\gamma}(p_1+k_2)_{\lambda}P_{\lambda\beta}(k_3)}{p_1^2k_2^2}]
\end{eqnarray}
   This addition corrects the vertex side of the Ward identity to have the massless poles of the inverse propagators on the right-hand side.  Of course,  none of these tweaks affect the UV properties.

This one-loop approach is guaranteed to be self-consistent in the UV, for reasons already given.  But it gives useful results in the IR too.  In fact, it reproduces an earlier suggestion \cite{corn138} for the 
  for the PT-RGI propagator:
\begin{eqnarray}
\label{fullprop}
\Delta^{-1}_{\alpha\nu}(p)  =  P_{\alpha\nu}(p^2)(p^2/g_0^2) & - & ~  \\ \nonumber
- \frac{11NP_{\alpha\nu}(p)}{48\pi^4}\int\!\mathrm{d}^4k\,\frac{1}{(k^2+m^2)((k+p)^2+m^2)}[p^2-\frac{m^2}{11}] & + & P_{\mu\nu}(p^2)M^2
\end{eqnarray}
where $M^2$ is chosen so that $\Delta$ has a pole at $p^2=-m^2$.
  This form is also essentially the same as given in \cite{corn076,corn090}.  As discussed in \cite{corn076,corn138}, a good approximation for  spacelike momentum is
\begin{equation}
\label{delta82}
\Delta(p)\equiv g^2(\mu )d(p) =  \frac{1}{(p^2+m^2)b\ln [\frac{p^2+4m^2}{\Lambda^2}]}.
\end{equation} 

From this we suggest that
\begin{equation}
\label{runningch}
\frac{1}{\bar{g}^2(q)}=b\ln [\frac{q^2+4m^2}{\Lambda^2}]
\end{equation}
is a decent approximation; a better form useable for both spacelike and timelike momentum is given in \cite{corn138}.

Next we study a version of $\phi^3_6$ that shows very similar results.

\section{\label{phisec}  The main issues for the three-vertex SDE, illustrated in $\phi^3_6$}

\begin{figure}
\begin{center}
\includegraphics[width=4in]{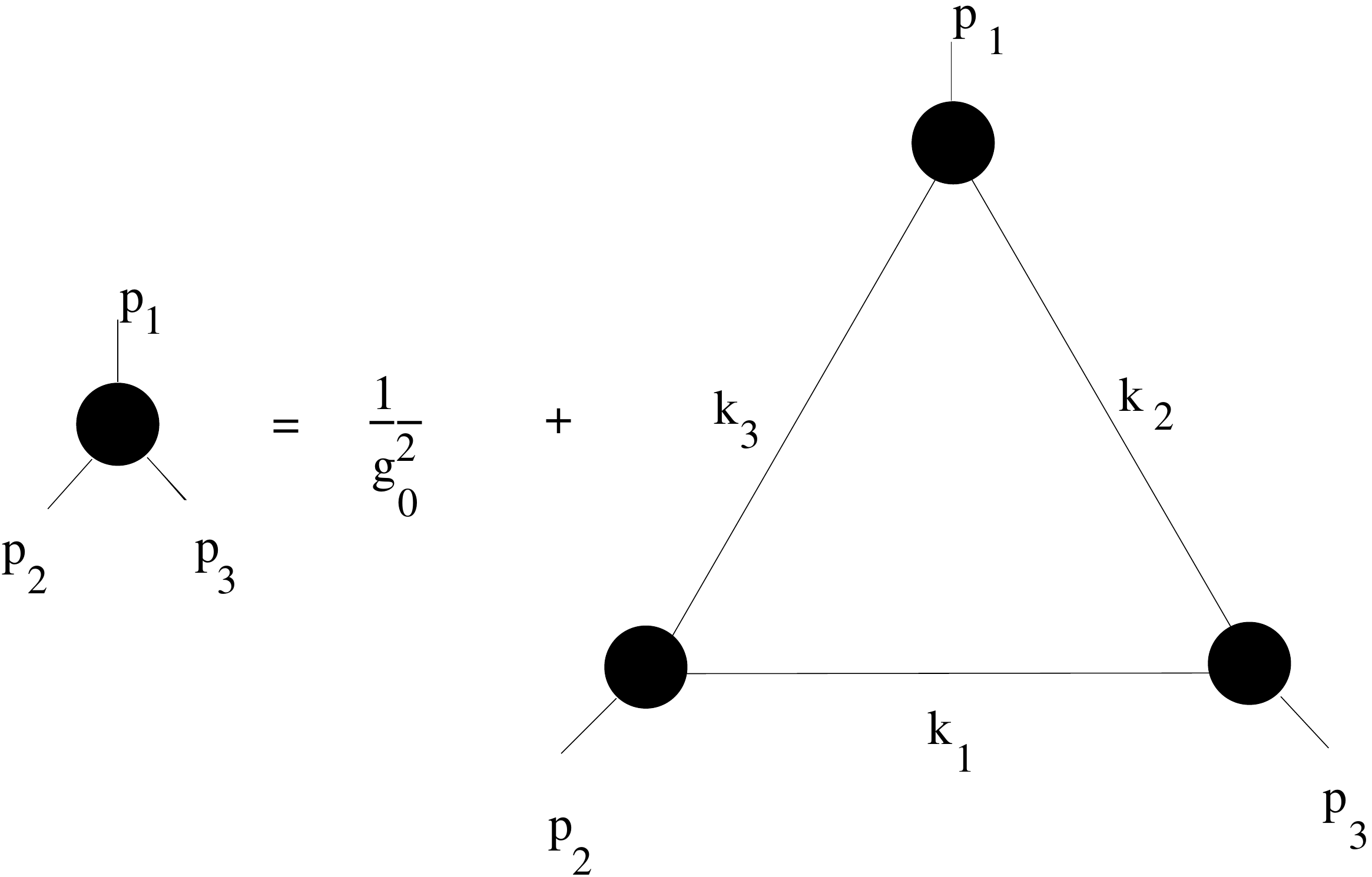}
\caption{\label{sde} The one-loop skeleton graph for $\phi^3_6$.}
\end{center}
\end{figure}

It is far too complicated to go very far with the SDEs for an NAGT, and now we will illustrate general features in a modified version of asymptotically-free $\phi^3_6$.   
Fig.~\ref{sde} shows  the only one-loop skeleton graph  for $\phi^3_6$.

Although there is no Ward identity as such in $\phi^3_6$, it is easy to construct one ``by hand", through giving two of the $\phi$ particles an Abelian charge.  We arrange coefficients so that the Ward identity for the charged particles is analogous to eqn.~(\ref{genwi}):
\begin{equation}
\label{phi3wi}
p_{1\alpha}G_{\alpha}(p_1,p_2,p_3)=\Delta^{-1}(p_2)-\Delta^{-1}(p_3).
\end{equation}
Analogously to eqn.~(\ref{vertequ}) we define a  scalar form factor $G(p_i)$ through:
\begin{equation}
\label{vertequ2}
G_{\alpha}(p_i)=(p_2+p_3)G(p_i)+\dots
\end{equation}
and equate this with the three-vertex of Fig.~\ref{sde}.
Then in the UV, where $p_i^2$ scales like a common large momentum $p^2$, it should be that that $G\Delta\rightarrow 1/p^2$.  We assume it is a decent approximation, therefore, to include IR effects by the simple rule
\begin{equation}
\label{simplerule}
G(p_i)\Delta (p) \rightarrow \frac{1}{p^2+m^2}.
\end{equation}
In effect, the vertex and propagator loop corrections cancel each other.  Then we assert that it is a good approximation, exact in the UV, to calculate the skeleton graph of Fig.~\ref{sde} by using free vertices and the free massive propagator of eqn.~(\ref{simplerule}).
The  result is:
\begin{equation}
\label{approxvert}
G(p_i)= \frac{1}{g_0^2}-b\int\![\mathrm{d}z]\ln [\frac{\Lambda_{UV}^2}{D+m^2}]
\end{equation} 
where, as in the gauge theory, at one loop the bare coupling is
\begin{equation}
\label{barecoupl}
\frac{1}{g_0^2}=b\ln (\frac{\Lambda_{UV}^2}{\Lambda^2})
\end{equation}
and
\begin{eqnarray}
\label{zint}
\int\![\mathrm{d}z] & = & 2\int_0\!\mathrm{d}z_1\,\int_0\!\mathrm{d}z_2\,\int_0\!\mathrm{d}z_3
 \,\delta (1-\sum z_i),\\
D & = & p_1^2\,z_2z_3+p_2^2\,z_3z_1+p_3^2\,z_1z_2 .
\end{eqnarray}
(The Feynman parameter $z_i$ goes with the line labeled $k_i$.) 
Clearly when any momentum is large, say $p_1^2\approx p^2 \gg m^2$, $G$ behaves like
\begin{equation}
\label{approxvert2}
G(p_i)= b\int\![\mathrm{d}z]\ln [\frac{D+m^2}{\Lambda^2}]\approx b\ln p^2.
\end{equation}
In the same UV limit we expect $\Delta \rightarrow 1/(bp^2\ln p^2)$ ({\em e.g.}, \cite{corn076}).  So the UV behavior, for large $k_i$, is self-consistent, since $G\Delta \approx 1/p^2$ and the integral corresponding to Fig.~\ref{sde} behaves exactly like our hypothesis.

In an NAGT a similar thing happens for $\Gamma^{(4)}$, as one sees by inspecting graphs and using PT-RGI Ward identities based on eqn.~(\ref{wardiden4}).  
For example, the one-loop skeleton graph with two $G^{(4)}$  depends only on the product $G^{(4)}\Delta \sim 1/p^2$, so as with the three-vertex we get an output vertex $\sim \ln p^2$ by inputting {\bf free} $G^{(4)},\Delta$.

\section{Running charge from the three-vertex:  A low-energy Ward identity}

The main point of this paper is the derivation of a ``low-energy" theorem for PT-RGI Green's functions that yields the running charge $\bar{g}^2(q)$, at all momenta, as the value of the scalar coefficient $G(0,q,-q)$ associated with Born-term kinematics in the three-vertex, evaluated when one momentum is zero.  The general Green's function $G(p_1,p_2,p_3)$ is in some sense the extension of the running charge to a gauge-invariant, scheme-independent, process-independent, renormalization-point independent vertex that depends on three momenta.

The running charge is related to the propagator via eqns.~(\ref{hdefinition},\ref{hruneqn}), or:
\begin{equation}
\label{zeromom2}
\Delta_{\beta\gamma}^{-1}(q)  = \frac{(q^2+m^2)}{\bar{g}^2(q)}P_{\beta\gamma}(q)+ \mathrm{gauge-fixing \;term}
\end{equation}
which has a massless longitudinal pole with residue $\sim m^2$.  (We need not indicate the momentum dependence of the mass.)

We are only interested in that part of the three-vertex having the Born kinematic structure, of the many different ones in this vertex.  Terms not having this structure include the longitudinal poles, so we can ignore the $m^2$ terms in eqn.~(\ref{zeromom}).
Saving only the kinematical structure of eqn.~(\ref{bornvert}), the linear terms both on the left and right of the Ward identity have the kinematics
\begin{equation}
\label{linterms}
p_{\alpha}[2q_{\alpha}\delta_{\beta\gamma}-\delta_{\alpha\beta}q_{\gamma}-\delta_{\alpha\gamma}q_{\beta}].
\end{equation}
Equating coefficients of the linear terms in the Ward identity at $p=0$ yields:
\begin{equation}
\label{lethm}
G(0,q,-q)=\bar{g}^{-2}(q).
\end{equation}
If  eqn.~(\ref{approxvert2}) holds for $G$, then
\begin{equation}
\label{approxrunch}
\bar{g}^{-2}(q)=b\int_0^1\!\mathrm{d}z\,2(1-z)\ln [\frac{q^2z(1-z)+m^2}{\Lambda^2}]
\end{equation}
which has the correct UV behavior, a threshhold at $-q^2=4m^2$, and a somewhat larger value of $\bar{g}^2(0)$ at the same value of the mass $m$ than given by the supposition of eqn.~(\ref{runningch}).  Such a discrepancy is only to be expected, given our approximations in the IR.  Probably the most accurate IR approximation to date is  that of \cite{corn138}, based on eqn.~(\ref{fullprop}) and other considerations that we cannot describe here.

The ``standard" RGI physical Green's function $\tilde{G}(p_1,p_2,p_3)$ of Eq.~(42) of \cite{corn138} is just: 
\begin{equation}
\label{tildevert}
\tilde{G}(p_1,p_2,p_3)= \frac{g\Gamma (p_1,p_2,p_3)}{[\tilde{Z}(p_1)\tilde{Z}(p_2)\tilde{Z}(p_3)]^{1/2}}=G(p_1,p_2,p_3)\bar{g}(p_1)\bar{g}(p_2)\bar{g}(p_3).
\end{equation}
In view of (\ref{lethm}) we have
\begin{equation}
\label{vertzero2}
\tilde{G}_{\alpha\beta\gamma}(0,q,-q)\approx G^0_{\alpha\beta\gamma}(0,q,-q)\bar{g}(0)+ \dots
\end{equation}
where the omitted terms have massless longitudinal poles, but do not contribute to the S-matrix.

[One must be careful to analyze the massless poles in the vertex and propagator before blindly using the Ward identity, because it is perfectly possible for there to be a pole in the vertex whose existence is unsuspected from the Ward identity alone.  For example, consider an Abelian vertex:
\begin{equation}
\label{poleex}
\Gamma_{\alpha}(p,q)=\frac{q_{\alpha}p\cdot q}{q^2}+\dots
\end{equation}
Then $q\cdot \Gamma =p\cdot q+\dots$ could as well have come from a term $\Gamma_{\alpha}=p_{\alpha}+\dots$ with no pole.  The difference of these two possible vertices is $p_{\beta}P_{\alpha\beta}(q)$.]

\section{The PT-RGI three-vertex and the beta-function}

Long ago \cite{corn099} there were speculations on deriving the beta-function from the PT three-vertex.  We can now be more precise.  From eqn.~(\ref{lethm}) and the usual definition of the beta-function equation for the running charge we find
\begin{equation}
\label{betaeqn}
\beta (\bar{g})=-\frac{\bar{g}^3}{2}\frac{q\partial}{\partial q}G(0,q,-q) 
\end{equation}
where on the left-hand side we could replace $\bar{g}=G(0,q,-q)$, but this is unnecessary.  If we use the speculation of eqn.~(\ref{approxrunch}) then the one-dressed-loop beta-function  is
\begin{equation}
\label{specbeta}
\beta (\bar{g})=-b\bar{g}^3\int_0^1\!\mathrm{d}z\,2(1-z)\frac{q^2z(1-z)}{q^2z(1-z)+m^2}
\end{equation}
from which we are to eliminate $q^2$ in favor of $\bar{g}^2$ with the  help of eqn.~(\ref{approxrunch}).  In perturbation theory ($m^2=0$) all of this trivially yields $\beta (g)=-bg^3$.  When $m^2\neq 0$ the elimination of $q^2$ in favor of $\bar{g}^2$ is not possible analytically, but the approximate running charge of eqn.~(\ref{runningch}) easily yields, with this procedure,
\begin{equation}
\label{massbeta}
\beta (g) = -bg^3[1-\frac{4m^2}{\Lambda^2}e^{-1/bg^2}],
\end{equation}
showing the expected non-perturbative behavior coming from $m^2$.

In general, a non-perturbative term such as the mass in the approximate running charge of eqn.~(\ref{runningch}), or the $\exp [-1/(bg^2)]$ in the beta-function, yields inverse powers of $q^2$ (modulo logarithms) in the UV asymptotics.  If the mass does not run, possibly corresponding to a bare mass term, this condensate could be interpreted as an $\langle A^2\rangle$ condensate, but this is not what we have in mind.  In conventional QCD the one-loop operator-product expansion of the PT propagator gives \cite{lavelle} a mass running in the UV as $m^2(q)\rightarrow C\langle G_{\mu\nu}^2\rangle/q^2$ where $C$ is a positive constant, given in \cite{lavelle}.   This, along with eqn.~(\ref{runningch}),  yields a condensate term in the UV expansion of the form
\begin{equation}
\label{condterm}
\bar{g}^{-2}(q)\rightarrow b\ln (\frac{q^2}{\Lambda^2})+\frac{4bC\langle G_{\mu\nu}^2\rangle}{q^4}+\dots
\end{equation}
  The fate of higher-order terms is not known, since this behavior is based on  a one-dressed-loop result.  Earlier it was suggested \cite{corn112} that condensate terms were closely-associated with the taming of the factorial divergences coming from IR renormalons; of course, a mass automatically does this.
\newpage

\appendix
\section{Conjectures on higher orders}

In the PT-RGI scheme, where the coupling constant automatically disappears from all skeleton expansions and SDEs, it is  not clear how to order these equations or even whether there is any ordering.  Using the 3-vertex in modified $\phi^3_6$ as an example, we propose here that the obvious ordering, by the number of loops in 3-vertex skeleton graphs, is useful, and contrast it to the usual perturbative ordering.  Let us call this scheme PT-RGI ordering.  We conjecture that in PT-RGI ordering all UV divergences are captured in the one- and two-loop skeleton graphs only, all others being UV-finite.  This may remind one of 't Hooft's remark \cite{thooft} that only the first two coefficients of the perturbative beta-function are scheme-independent, and that it is actually possible to choose a scheme where all other coefficients vanish.  We emphasize that this does not mean that the running charge, or other quantities normally extracted from the beta function, is given exactly by saving only two coefficients in the beta function.  In fact, in the PT-RGI scheme the beta function is at best a derived quantity that can be read off from the running charge as calculated (uniquely) from the SDEs of this scheme.  All-order effects are still critical, and contain, for example, the contribution of solitons and condensates \cite{corn112}.

Our conjecture seems fairly firm for modified $\phi^3_6$, and we extend it to NAGTs.  However, here there are immense complications from vertex numerators, and everything we say for higher orders is subject to rebuttal.  Nonetheless, we are not aware of any potential contradictions to the higher-order PT-RGI conjecture.

Rather than give a general proof of the higher-order behavior for NAGTs, which would be extemely lengthy, we illustrate the general principles with some selected two-loop skeleton graphs.  These principles apply to all orders in $\phi^3_6$.  The main point is that the number of loops in any skeleton graph for the 3-vertex is in direct correspondence with the leading UV behavior of the exact vertex, in such a way that the leading UV behavior of the vertex is exactly given by the one-loop skeleton graph.    The skeleton-graph expansion for NAGTs is complicated; further details for NAGTs, many but not all of which hold for the PT, are given in \cite{blee}. As pointed out there, the skeleton expansion we and they use is free of overlapping divergences, and can be subtractively renormalized.

       Consider first any $\phi^3_6$ skeleton graph in which all vertices and propagators are replaced by bare vertices and propagators.  In the ``by-hand" version of one-loop skeleton graphs for $\phi^3_6$ used in Sec.~\ref{phisec},   which mimics the PT Ward identities, we have already explained that with these identities the exact leading UV behavior comes from this replacement.  This is not true for skeleton graphs with more than one loop, but it is still a useful starting point.   All graphs are defined by first combining both loop momenta with Feynman parameters, introducing a UV cutoff $\Lambda_{UV}$, and integrating over momenta first.  
The result for any $N$-loop skeleton graph $G_N$ is of the form (irrelevant factors omitted) with exactly one UV logarithm after the momentum integrals.  In interpreting this logarithm, recall that the vertex $G$ actually represents an inverse running charge.  We have:
\begin{equation}
G_N=\int\!\prod \mathrm{d}z_i\delta (1-\sum z_i)U^{-3}\ln \left[\frac{\varphi /U+ m^2}{\Lambda_{UV}^2}\right]
\end{equation}
where $U$ is the determinant of the graph and $\varphi$ is a sum of positive monomials in the $z_i$ multiplied by scalar products of external momenta, such as $k_a\cdot k_b$.    An important property of $U$, a sum of $N^{th}$-order positive monomials in the $z_i$, is that if all the parameters $z_i$ of a single loop are scaled through $z_i\rightarrow \lambda z_i$, $U$ scales linearly in $\lambda$.  But if less than all the lines are scaled, $U$ remains finite in the limit $\lambda\rightarrow$ 0 for generic values of the other Feynman parameters.  Any further UV divergences, if any, introduced by integrals over the Feynman parameters must therefore come from scaling all the lines of one loop toward zero.  But because the graph is a skeleton graph there are no such divergences, as we now review.
It is easy to see that all skeleton graphs for the vertex are, in the language of graph theorists, of girth four \cite{corn112}, meaning that any internal loop has at least four lines. This follows simply from the lack of vertex insertions in the skeleton graphs.   Take any loop of girth $K$ and scale the $K$ Feynman parameters of this loop by a factor $\lambda$. If there is any singularity in the Feynman-parameter integral it will show up as a divergence for small $\lambda$.  All the relevant factors involving $\lambda$, when this variable is small, are:
\begin{equation}
\label{cornmor}
\int\!\mathrm{d}\lambda\,\frac{\lambda^{K-1}}{\lambda^3} 
\end{equation}
where the denominator comes from $U^3$.  Since $K >$ 3, this integral cannot diverge.  A similar argument holds when the parameters of more than one loop are simultaneously scaled.  Consequently, after renormalization any skeleton graph with bare vertices and propagators has exactly one UV logarithm.

Now dress all the vertices and propagators.  An $N$-loop skeleton graph has $N_G=2N+1$ vertices and $N_{\Delta}=3N$ propagators.  The exact UV behavior of any vertex $G$ in the graph is $G\sim \ln k^2$, and similarly $\Delta^{-1} \sim \ln k^2$.  Moreover, all skeleton graphs for the vertex are dimensionless.  Therefore, {\em schematically} the $N^{th}$-order term of the vertex looks like:
\begin{equation}
\label{intbeh}
G_N=[\frac{1}{g_0^2}]_N - const. \int^{\Lambda_{UV}}\!\mathrm{d}k^2\frac{1}{[k^2+p^2][\ln (k^2+p^2)]^{N-1}}
\end{equation}
where $p$ is a generic external momentum and $[1/g_0^2]_N$ is an $N$-loop contribution to the bare coupling.  

\begin{figure}
\begin{center}
\includegraphics[width=2in]{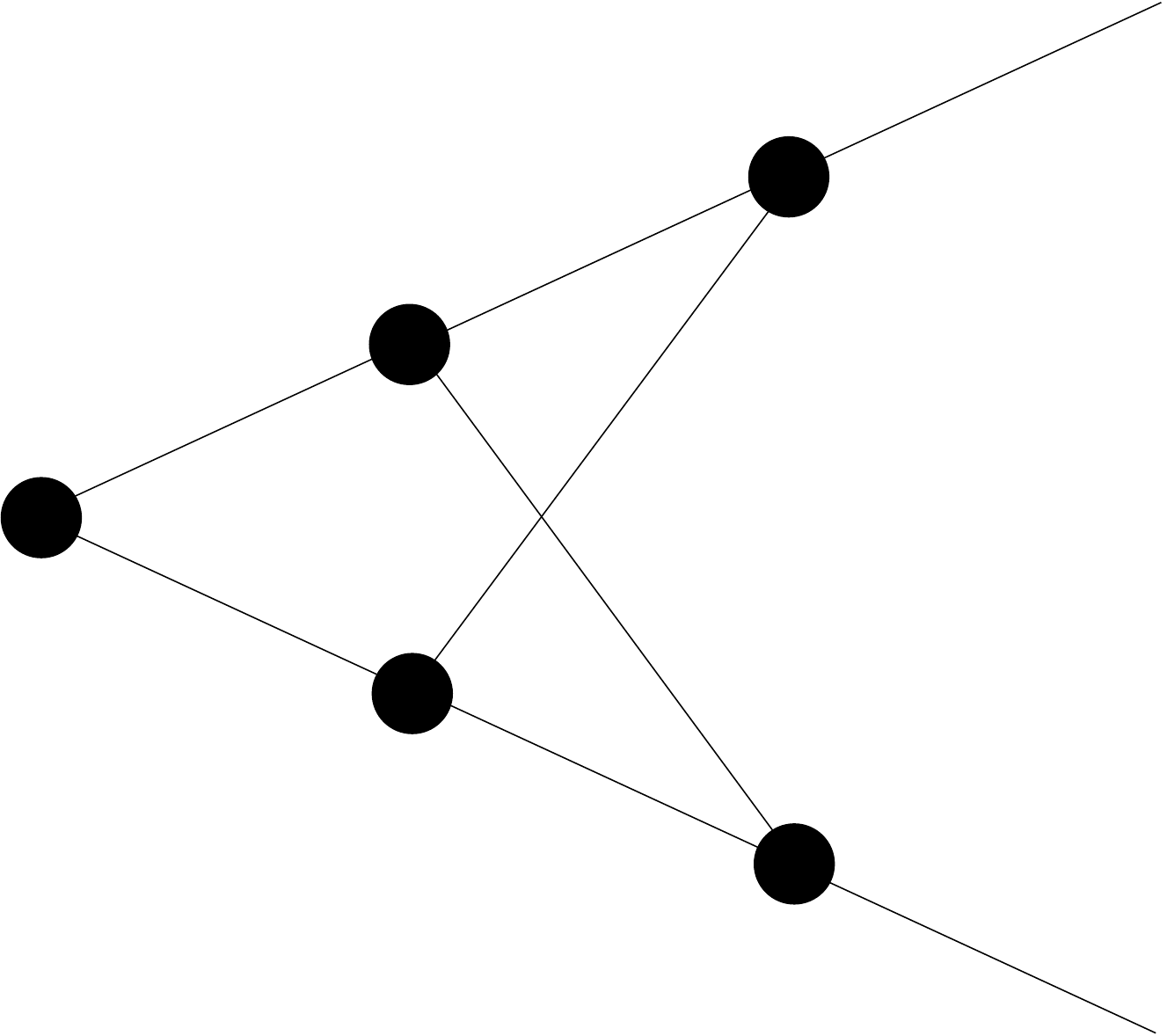}
\caption{\label{2pi} The two-loop 2PI skeleton graph for   the   three-gluon proper vertex, divided by $g^2$.  Black circles are the vertex $G$ and lines are the propagator $\Delta$.   }
\end{center}
\end{figure}

Consider now $N=2$, for which  the only skeleton graph is
  illustrated in Fig.~\ref{2pi}.
At $N$=2 the integration in Eq.~(\ref{intbeh}) gives 
\begin{equation}
\label{intbeh2}
-\ln [\ln \Lambda_{UV}^2]+\ln [\ln p^2]
\end{equation} 
which A) has the correct momentum dependence for a two-loop contribution to the inverse running charge $\bar{g}^{-2}(p^2)$; B) can be additively renormalized from a term in $[1/g_0^2]_2$ behaving like $\ln [\ln \Lambda_{UV}^2]$.  For $N>2$ the integral is cutoff-independent, and has a large-$p$ dependence $[\ln p^2]^{2-N}$ (modulo terms less singular than a logarithm).  

Of course, the actual two-loop integral is much more complicated.  We have studied  these complications to some extent, and believe that they do not alter the simple conclusions drawn from the schematic form of Eq.~(\ref{intbeh}).

\begin{figure}
\begin{center}
\includegraphics[width=2in]{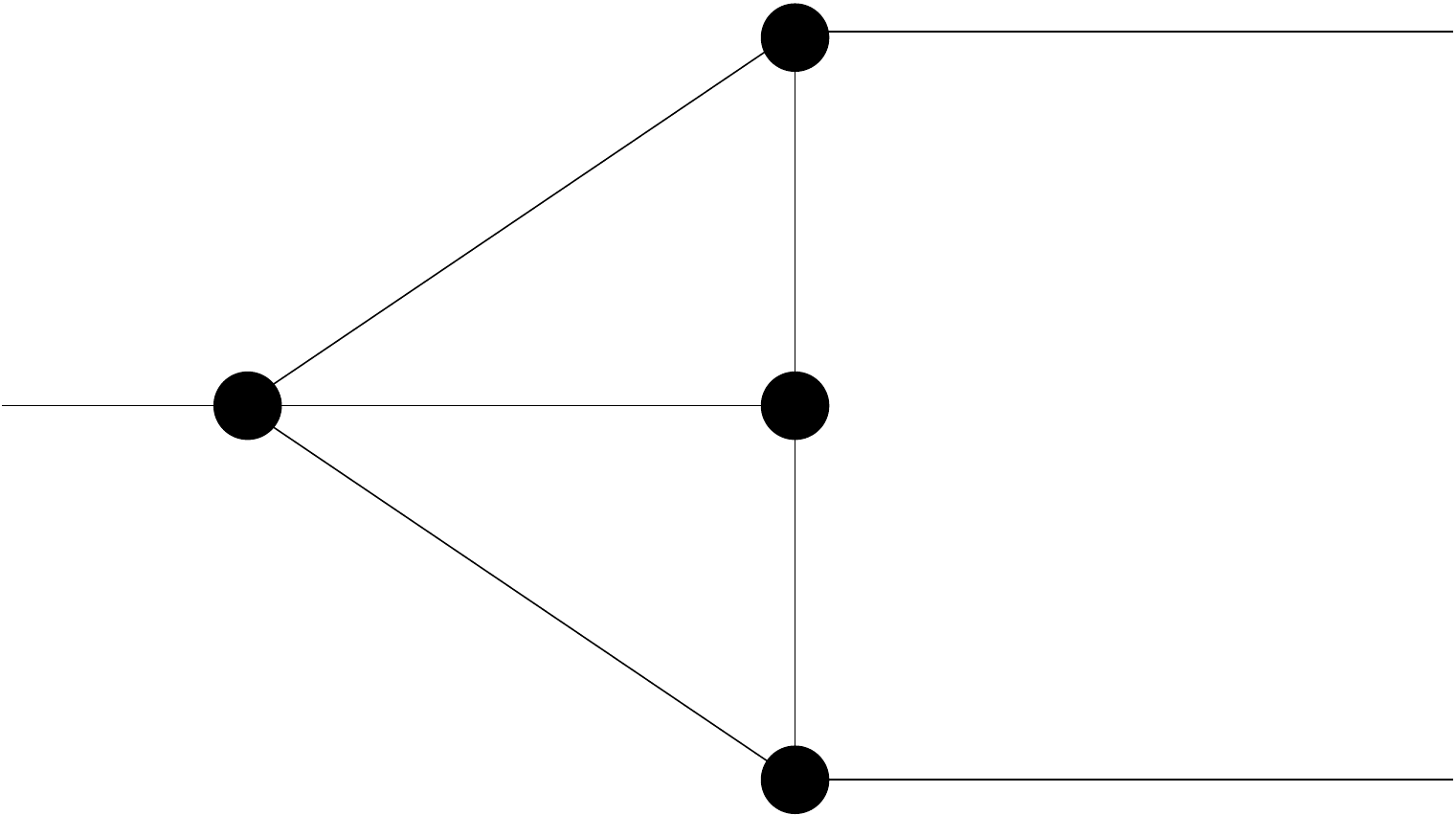}
\caption{\label{2pi-2} The two-loop 2PI skeleton graph with a four-point vertex, for   the   three-gluon proper vertex, divided by $g^2$.  Black circles are the vertex $G$ and lines are the propagator $\Delta$.   }
\end{center}
\end{figure}

For NAGTs the argument must be modified.  Graphs with 4-point vertices, such as Fig.~\ref{2pi-2}, have girth $K$=3.  If it were not for numerator factors this would not change things, because in $d=4$ the determinant $U$ appears to the power -2 (generically -$d/2$), so 
\begin{equation}
\label{2loop4}
\int\!\mathrm{d}\lambda\,\frac{\lambda^{K-1}}{\lambda^2}
\end{equation}
is finite for small $\lambda$ at $K$=3.  It appears that the momentum-space integration produces only a single logarithm, because vertex numerators effectively change the graphs to those of a $d=6$ scalar theory.  This assertion, however, is extremely difficult to prove.

\section{UV asymptotic behavior from the beta function}

Here we review what should be a well-known argument that in an asymptotically-free theory the RGI equation for the running charge $\bar{g}^2(p)$ in a conventional scheme for the beta function, truncated at any finite order, however large, leads to an inverse running charge $\bar{g}^{-2}(p)$ whose UV growth at sufficiently-large $p$ is uniquely determined by the first two terms of the beta function (which themselves are scheme-independent).  The remainder terms vanish asymptotically.  A generic asymptotically-free beta function shows factorial growth, and truncating it   at a finite order $J$, however large, is delicate, but that of course is what one must do in practice. Factorial growth means that  the regime of sufficiently-large momentum is bounded below by a momentum growing    with $J$  at an exponential rate.  

The equation for the running charge is: 
\begin{equation}
\label{runch}
p^2\frac{\partial}{\partial p^2}\bar{g}^2(p^2)=\beta (\bar{g})=-\sum_0 b_j\bar{g}^{2j+4}
\end{equation}
where $b_0\equiv b$, the usual one-loop coefficient.  Introduce the variables 
\begin{equation}
\label{newvar}
\bar{g}^2=1/L,\;\;t= \ln (p^2/\Lambda^2)
\end{equation}
where $\Lambda$ is the physical QCD mass scale.  Equation (\ref{runch}) becomes
\begin{equation}
\label{newrunch}
\dot{L}=\sum \frac{b_j}{L^j}
\end{equation}
where the dot means a $t$ derivative.
To make contact with perturbation theory, we truncate the sum over $j$ at a maximum value $J$, and restrict momenta to a regime in which $L\gg (b_j)^{1/j}$ for all $j\leq J$.  The existence of such a regime will become apparent later. Then (measuring momenta in units of $\Lambda$)
\begin{equation}
\label{runchsol}
b\ln p^2 =\int\!\mathrm{d}L\;L^J\frac{1}{L^J+b^{-1}\sum_1^J b_jL^{J-j}}.
\end{equation}
Decompose the integrand as:
\begin{equation}
\label{decom}
L^J\frac{1}{L^J+b^{-1}\sum_1^J b_jL^{J-j}}=1+\sum \frac{\alpha_j}{L+\beta_j}.
\end{equation}
For $J>4$ these equations have no explicit solution, but all we need to know is that for any $J$
\begin{equation}
\label{decom2}
\sum \alpha_j=\frac{-b_1}{b}.
\end{equation}
Let us define the UV region of interest as the region where $L\gg Max[\beta_j]$.  In this region we can drop the $\beta_j$ in the numerator and come to:
\begin{equation}
\label{betasol}
L= b\ln p^2+\frac{b_1}{b}\ln L+\mathcal{O}(\frac{1}{L})
\end{equation}
showing that the inverse running charge can be approximated arbitrarily well at sufficiently large momenta using only the first two terms of the beta function truncated at some $J$, whose value depends on momentum in general. 

This cannot work for the untruncated beta function, which we know has factorial growth and hence renormalons.  However, it has been argued for $\phi^3_6$ \cite{corn112} that mass generation (an IR effect) effectively removes the Borel singularities from the factorial growth, and introduces condensate terms vanishing like an inverse integral power of $p^2$.

\end{document}